%% file: main.tex
\documentclass[sigconf,pbalance]{acmart}

\usepackage{subcaption}
\usepackage{calc}
\usepackage{amstext}
\usepackage{amsmath}
\usepackage{amsthm}
\usepackage{multicol}
\usepackage{pslatex}
\usepackage{algorithm2e}
\usepackage[bottom]{footmisc}
\usepackage{todonotes}
\setuptodonotes{inline}
\usepackage{url}
\usepackage{graphicx}
\usepackage{booktabs}

\newcommand{\wifi}{\mbox{Wi-Fi}}
\newcommand{\piso}{Piso-WiFi}

\setcopyright{acmlicensed}
\copyrightyear{2026}
\acmYear{2026}
\acmDOI{XXXXXXX.XXXXXXX}
\acmConference[SPAIC'26]{1st Security and Privacy for Asian Internet Communities, AsiaCCS Workshops}{June 02, 2026}{Bangalore, India}
\acmISBN{978-1-4503-XXXX-X/2018/06}

\begin{document}

\title[Open Challenges for Secure and Scalable Wi-Fi Connectivity in Rural Areas]{Open Challenges for Secure and Scalable\\Wi-Fi Connectivity in Rural Areas}

\author{Philip Virgil Berrer Astillo}
\affiliation{%
  \institution{University of San Carlos, Philippines}
  \country{}
}
\email{pvbastillo@usc.edu.ph}

\author{Jayasree Sengupta}
\orcid{0000-0001-7519-5293}
\affiliation{%
  \institution{IIIT Allahabad, India}
  \country{}}
\email{jayasree@iiita.ac.in}

\author{Mathy Vanhoef}
\affiliation{%
  \institution{DistriNet, KU Leuven, Belgium}
  \country{}
}
\email{mathy.vanhoef@kuleuven.be}

\begin{abstract}


Providing reliable, affordable, and secure Internet connectivity in rural areas remains a major challenge. Pay-for-use Wi-Fi hotspots are emerging as a scalable solution to provide affordable Internet access in underserved and rural regions. Despite their growing adoption, their security properties remain largely unexplored. In this paper, we present a security analysis of these hotspot ecosystems based on Wi-Fi surveys and practical attack validation. We first perform a \wifi{} survey conducted in two countries, namely the Philippines and India, to understand the deployment and adoption of such systems in practice. Our results suggest that \piso{} pay-to-use hotspots are particularly widespread in rural regions of the Philippines, and that India’s PM-WANI initiative is slowly gaining traction. We then perform a security assessment of these deployments and demonstrate two practical attacks: (1) hijacking another user’s paid connection; and (2) rogue hotspots. We analyze the root causes of these vulnerabilities, introduce threat models tailored to pay-for-use hotspot deployments, and outline practical security improvements, including a secure caching architecture. Our findings highlight security challenges in emerging rural connectivity infrastructure and provide directions toward more secure and scalable deployments.
\end{abstract}


\maketitle

\section{Introduction}
\label{sec:introduction}

Reliable and affordable Internet connectivity remains a major challenge in many developing regions. While urban areas increasingly have widespread broadband coverage, rural communities often rely on limited and expensive data services.
Furthermore, cellular networks often provide limited bandwidth and can incur substantial costs as data usage increases~\cite{itu2023}.
This limited Internet access causes a digital divide, restricting access to information, education, and economic opportunities.
As a result, providing affordable, reliable, and secure Internet access in rural areas remains an important challenge.

One promising approach to increase Internet connectivity is the emergence of community-operated pay-for-use \wifi{} hotspots. These act as a scalable solution to provide affordable Internet connectivity in underserved and rural regions.
Such systems allow anyone with connectivity to act as a micro–service provider, thereby offering community-based Internet coverage without requiring large-scale infrastructure investments.
These systems are typically deployed using low-cost routers, making them especially attractive in rural areas where broadband and cellular Internet are limited~\cite{gupta2023-pmwani}. 

Despite their growing adoption, the security properties of these pay-for-use hotspot ecosystems remain largely unexplored. Unlike traditional enterprise or managed public Wi-Fi deployments, these systems operate in decentralized environments, often using commodity hardware and minimal configuration. This creates unique security challenges that differ from traditional Wi-Fi deployments. Understanding the security risks of these emerging connectivity models is useful in order to ensure safe and sustainable Internet access in developing regions.

To better understand the adoption of these systems, we conducted a \wifi{} survey in two countries: the Philippines and India.
Our observations suggest that \piso{} coin-operated hotspots, named after the Philippines' peso currency, are widely deployed, particularly in rural and semi-urban regions, while India’s PM-WANI (Prime Minister’s \wifi{} Access Network Interface) initiative is slowly gaining traction.
Unlike \piso{}, PM-WANI is not coin-operated, but relies on online subscriptions. These observations provide an initial view into how pay-for-use \wifi{} can be a growing solution to provide affordable Internet connectivity, especially in rural regions.

Motivated by these observations, we perform a security review of \piso{}.
This showed that \piso{} is vulnerable to known attack techniques that allow an adversary to (1) hijack another user’s active session, and (2) set up a rogue \piso{} hotspot to impersonate legitimate providers.
Both attacks enable an adversary to obtain free Internet access.
We implemented and validated both attacks in real-world settings, confirming that they can be exploited with commodity hardware and minimal technical expertise.
Next, we analyze India’s PM-WANI framework, which is another pay-for-use \wifi{} hotspot model.
It uses a different payment system with an online subscription. Based on its design and publicly available specifications, we identify potential security vulnerabilities that may arise in similar ways to those observed in \piso{}. However, we emphasize that this analysis is preliminary and does not include a practical attack evaluation.

Lastly, we discuss these common challenges and outline open problems in designing secure, low-cost connectivity mechanisms for rural regions.
Notably, we outline two threat models that can be used as a basis for better securing pay-to-use hotspots, and discuss secure local caching approaches that can help optimize limited backhaul bandwidth while preserving security.
Overall, our exploratory study shows that more research is needed to provide secure but easy-to-use and scalable public \wifi{} hotspots.


Summarized, our contributions are:
\begin{itemize}
    \item We perform a systematic \wifi{} survey in the Philippines and India, study security trends, and discuss the usage of pay-to-use \wifi{} hotspots (§~\ref{sec:wardrive}).
    \item We conduct a security assessment of \piso{} and demonstrate the feasibility of two attacks: hijacking connections and rogue hotspots (§~\ref{sec:piso}).
    \item We conduct a preliminary security assessment of the PM-WANI hotspots, and compare to \piso{} (§~\ref{sec:pm-wani}).
    \item We define new threat models for pay-to-use hotspots and propose ways to improve their security (§~\ref{sec:threats}).
\end{itemize}
Additionally, we cover related work in §~\ref{sec:relatedwork}, and conclude in §~\ref{sec:conclusion}.

\section{Background}
\label{sec:background}

In this section, we introduce the IEEE 802.11 standard that underpins \wifi{}, explain how clients can discover nearby networks, and cover security and authentication aspects of 802.11.

\subsection{Network discovery}

When a \wifi{} client wants to connect to a wireless network, it first discovers nearby Access Points~(APs) through either passive or active scanning.
In passive scanning, the station listens for beacon frames periodically broadcast by nearby APs. Each beacon advertises the network's name, also called the Service Set Identifier (SSID), as well as other network properties such as supported data rates, security capabilities, and other configuration parameters.
In active scanning, the station transmits broadcast probe request frames.
Nearby APs reply with probe responses containing similar information to that found in beacons.
Both mechanisms allow clients to learn which networks are within their range and to decide which AP to associate with, typically based on received signal strength and preferred SSIDs.

Each individual AP has a unique Basic Service Set~(BSS) Identifier, which typically equals the MAC address of the AP.
Multiple APs can advertise the same network name, i.e., the same SSID.

\subsection{Security of IEEE 802.11}

Once an AP is selected, the client begins the authentication and association stages.
During the authentication stage, the client performs open system authentication, which involves a simple exchange of request and response frames without actual credential validation.
Only when using WPA3-Personal does actual authentication take place at this stage of the connection process~\cite{ieee80211-2024}.
When using WPA1/2 or WPA3-Enterprise, actual authentication takes place after association.
Next, regardless of which WPA version or variant is used, the \mbox{4-way} handshake is performed to negotiate session keys to protect and encrypt data frames.

Unicast \wifi{} frames between the client and the AP are protected using pairwise session keys, and group-addressed frames, i.e., broadcast or multicast frames, are protected using a group key.
This also implies that the cipher used to protect unicast and group-addressed frames can differ, i.e., networks can be configured using a different pairwise and group cipher, respectively.

Lastly, many commercial or community \wifi{} systems, such as public hotspots, employ captive portals after association.
These portals intercept HTTP traffic and redirect the user to a local payment or authentication webpage, or a page showing an end-user license agreement, before granting broader Internet access.
While a common solution in practice, captive portals are not part of the IEEE 802.11 standard itself, and their security is therefore not always well-integrated and implementation-dependent, as we later discuss.

\section{Wi-Fi Surveys}
\label{sec:wardrive}

In this section, we present the methodology used for \wifi{} surveys in rural areas of the Philippines and in India to study the prevalence of pay-to-use hotspots.

\subsection{Methodology}

To detect nearby \wifi{} networks, we used the WiGLE Android app\footnote{Available at \url{https://wigle.net/tools}} while driving around various cities in the Philippines and India.
This app uses Android's built-in functionality to detect nearby networks, meaning it will listen for beacon frames and will periodically trigger the transmission of broadcast probe requests.

During a \wifi{} survey, WiGLE will log each detected AP, and store the MAC address (BSSID) of the AP, the SSID of the network, the geolocation, and other metadata.
Importantly, WiGLE will also log the network security properties of the network as reported by Android's scan results.

More precisely, WiGLE takes the \verb|ScanResult| object, which describes information about a detected access~\cite{android-scanresult}, and logs the capability string that is provided by the \verb|capabilities| field of the scan result.
Although there is no documentation on how this string is constructed, we can inspect the source code of Android’s \wifi{} Information Element parser, and specifically the function \verb|generateCapabilitiesString|, to learn the information it provides.

First, the capabilities string contains information on whether the network is an infrastructure or ad-hoc network, whether Wi-Fi Protected Setup~(WPS) is supported by the network, and whether Management Frame Protection~(MFP) is either supported or required.
Second, the network's supported authentication methods are listed, e.g., WPA1/2 password authentication, EAP-based Enterprise authentication, WPA3 password authentication with SAE, opportunistic wireless encryption, along with supported key sizes.
Lastly, the capabilities also include the cipher suite used to protect and encrypt data once the station is authenticated with the network.

We remark that captive-portal authentication cannot be detected using passive \wifi{} surveys.
This is because these deployments do not use authentication at the \wifi{} layer, but instead intercept HTTP traffic, and perform authentication once the station is already connected to the \wifi{} network.
Additionally, the group cipher used to protect multicast and broadcast traffic is not included in the network's capabilities string, meaning it cannot be collected when using WiGLE.

\subsection{Philippines}

We performed a \wifi{} survey in the Philippines in eastern Mindanao, while driving around the rural areas and cities near the coastline.
One challenge is that rural areas have few \wifi{} networks, requiring us to perform surveys of several hours to detect a representative number of \wifi{} networks.
After driving around for three days in total, during which we drove for roughly 4 hours, we detected a total of 2\,620 \wifi{} networks.

\input{table-survey}

A first surprising observation is that close to 20\% of detected networks still support the old TKIP encryption cipher of WPA1 as the pairwise cipher (see Table~\ref{table:survey}).
This cipher has been deprecated by the \wifi{} Alliance and is known to be vulnerable to various side-channel attacks~\cite{vanhoef2013practical,schepers2019practical}.
A possible root cause of the continuing support for TKIP in practice is that people in rural areas may update their devices less frequently, causing many APs to still support this outdated cipher.
For comparison, in India roughly 13\% of networks still support TKIP as the pairwise cipher, and prior work reported TKIP support in 17.99\% to 54.23\% of networks across selected European countries in 2019~\cite{schepers2021let}.

To detect how many \piso{} hotspots there are, we can easily check if the string \verb|piso| occurs in the SSID.
This works because the main way for users to detect \piso{} hotspots is based on the SSID, hence why these hotspots practically always include the string \verb|piso|.
This revealed that close to 5\% of the networks in our dataset are identified as \piso{} hotspots, showing that they are very popular in the region we surveyed.

\subsection{India}

In India, we performed a \wifi{} survey using WiGLE in areas of Bangalore, Calcutta, and more rural areas near the outskirts of Allahabad.
This was done by surveying each city for roughly an hour, spread out over two days, detecting 6\,464 networks (see Table~\ref{table:survey}).

Unlike \piso{}, the PM-WANI hotspots cannot be detected based on their SSID, i.e., their network name.
Instead, there is an online list of the MAC addresses of PM-WANI access points.
Users of PM-WANI use mobile apps that can download the list and notify the user when a PM-WANI network is nearby.
To detect these hotspots ourselves, we scrape the list of MAC addresses of PM-WANI networks and intersect this with the set of detected networks in our \wifi{} survey.
This revealed that only 3 networks are PM-WANI hotspots in our dataset (see Table~\ref{table:survey}), which limits the conclusions that can be drawn about their broader deployment.

We believe the low number of PM-WANI hotspots compared to \piso{} is due to the lengthier process of setting up a PM-WANI hotspot: users must first register their business, and can only then set up a hotspot.
In contrast, a \piso{} hotspot in the Philippines requires no registration, but only the provisioning of a router.

Finally, we observed that around 13\% of networks still support the old and broken TKIP encryption cipher.
Although lower than our survey in the Philippines, this is still a worrying number of networks that are potentially vulnerable to attacks.

\subsection{Limitations of the Measurement Study}

Our Wi-Fi survey provides an initial snapshot of pay-for-use hotspot deployments. However, it has a couple of limitations. Data collection was opportunistic and conducted while driving through selected regions, which may introduce sampling bias. The observation period was also relatively short. In addition, the number of detected PM-WANI hotspots in our dataset is small, so observations regarding PM-WANI should be interpreted as preliminary. Despite this, our dataset offers useful qualitative insights into the deployment and security characteristics of emerging pay-for-use hotspot systems.

\section{Security of PISO-WIFI}
\label{sec:piso}

In this section, we analyse the security of \piso{}, and empirically confirm that it does not defend, or otherwise mitigate, two attacks that can hijack a user's paid Internet connection. All attacks described in this section were experimentally validated in controlled and real-world settings using our own devices.

\subsection{Background on \piso{}}

The idea behind \piso{} hotspots is that ordinary users can share their Internet connection to earn extra income by providing Internet access to nearby users.
To use the hotspot, nearby individuals can connect to the \wifi{} network, and then insert a 1 peso coin to be given 5-10 minutes of unlimited internet access.
Many individuals have embraced the idea of acquiring \piso{} Machines to have an additional source of income.
These machines are commonly installed near Sari-Sari Stores (small shops), along roadsides for use by passing travelers, and also near residential communities or near local cafe shops.
The average cost of a new machine
is approximately 9,000 pesos~\cite{piso-cost}.

As also indicated by our \wifi{} survey in the previous section, such networks are particularly common in more rural areas and barangays, where a significant portion of the population lacks the financial means to own their own Internet connection.
In essence, these machines provide an affordable opportunity to connect to the Internet, especially in rural areas.

\input{fig-piso}

Figure~\ref{fig:piso} shows how users connect to \piso{} hotspots.
First, users initiate the process by connecting to the \wifi{} hotspot to access the system's captive portal.
Subsequently, the system guides users through inserting coins into the physical \piso{} machine.
To make it easier to locate the machine, most will show blinking lights when coins need to be inserted.
After inserting coins, the users can click a button to finish the payment, after which the corresponding usage time limit is calculated and Internet access is granted.
This user-friendly design, with low overhead, is especially important since the target audience is non-technical people who may also not always own the most recent smartphones or IT equipment.

\subsection{User Masquerading Attack}
\label{sec:piso:masquerade}

In our first experiment, we test whether \piso{} tries to prevent or mitigate the known attack technique of spoofing the MAC address of an already-connected client.
Note that modern APs have features that attempt to detect such attacks~\cite{cisco-anomalous-detection}.
We test for these attacks in two phases: first in a controlled environment, and then under more real-world conditions.
Note that for ethical reasons, we only test the attack against our own devices, and never against real users.

\subsubsection{Controlled experiment}

We first tested whether a user's connection can be hijacked in a controlled environment.
In this environment, we assume the adversary knows in advance what the MAC address and IP address of the victim are, and we assume that the adversary knows exactly when the victim has paid for their \piso{} Internet connection.
Once the victim client has connected and paid, the adversary spoofs the MAC and IP address of the victim, and connects to the \piso{} access point.
In our experiments, this attack was always successful.
We did observe that, when the victim tries to reconnect, the Internet connection of the adversary is disrupted.
In other words, as long as the real user is present, their device might try to automatically reconnect, which can interfere with the adversary's hijacked \piso{} connection.
Nevertheless, once the victim goes out of range, or when they try to connect to a different \wifi{} network, the hijacked connection becomes stable.

\subsubsection{Real-world attack evaluation}

As a second experiment, we more accurately simulate a real attack, where the adversary does not in advance know the MAC address of the victim, and does not know when the victim will connect to \piso{}, i.e., we follow a \emph{blind testing} approach.
To handle this, the adversary will first use a \wifi{} monitoring tool such as \verb|airodump-ng| to determine which clients are currently connected to the \piso{} network.
Once all connected clients have been determined, the adversary can track how many data frames each client is exchanging with the hotspot.
A high number of exchanged data frames indicates that the client has successfully connected with \piso{} and is actively using the connection.
In other words, seeing many data frames likely means the client has paid for Internet access.
Since \piso{} does not use encryption at the \wifi{} layer, the adversary can trivially learn the MAC address and IP address that a victim client is using.
The adversary can then spoof this MAC and IP address, similar to our controlled experiment, to hijack the victim's paid Internet connection.
We successfully confirmed this attack strategy in practice.

\subsection{Rogue Access Point}
\label{sec:piso:rogue}

\input{fig-piso-mitm}

In case determining connected clients is challenging, e.g., due to the hidden station problem, or due to beamforming making it difficult to sniff the frames of clients~\cite{antonioli2017practical}, then an alternative attack strategy is creating a rogue clone of a \piso{} hotspot (see Figure~\ref{fig:rogue}).
The idea is that the adversary creates an almost identical clone of a real \piso{} network, which is possible using commodity off-the-shelf tools such as \verb|hostapd| on Linux.
The adversary can then either wait until a victim connects to the adversary's rogue AP, or can abuse techniques such as the spoofing of Channel Switch Announcements to force the victim into connecting with the rogue AP~\cite{vanhoef2020protecting}.
The attacker can then conduct a
Man-in-the-Middle~(MitM) Attack, by positioning themselves
between the client device and a real \piso{} network, allowing
them to intercept and manipulate communication between
the two parties.
Concretely, the victim is tricked into paying for the adversary's connection to the \piso{} hotspot, after which the adversary can freely access the Internet.

\section{India's PM-WANI}
\label{sec:pm-wani}

In this section, we give a description of how the PM-WANI hotspots in India work, and compare it with the \piso{} hotspots of the Philippines (see Table \ref{tab:pmwani-piso}).

\subsection{Overview of PM-WANI}

India’s growing digital economy has made reliable broadband Internet access ever more important.
However, although access to mobile Internet in India has increased significantly in recent years, rural and remote regions often still lack affordable and high-quality Internet access.
To address these challenges, and provide affordable Internet access to all citizens, India's Department of Telecommunications approved the Prime Minister’s \wifi{} Access Network Interface (PM-WANI) framework in 2020 \cite{dot2020pmwani}.

Compared to \piso{}, the PM-WANI system is a more centrally managed but still \emph{federated} architecture, with a similar goal of providing Internet access to all regions in the country.

In this federated model, responsibilities or various tasks are split between the following stakeholders:

\begin{itemize}
    \item \textbf{Public Data Office (PDO):} Will establish and operate compliant \wifi{} hotspots that provide Internet access.
    \item \textbf{Public Data Office Aggregator (PDOA):} They will aggregate PDOs, including the PDOs' \wifi{} APs, network names, and locations. The PDOA also has the responsibility of performing authentication and accounting of users wanting to use a hotspot, handling payment transactions, and keeping track of each user's usage. PDOAs must register with the government.
    \item \textbf{App Providers:} Will create mobile apps that allow users to discover nearby PM-WANI \wifi{} hotspots. This implies that users must have such an app installed in order to use PM-WANI. App providers must register with the government.
    \item \textbf{Central Registry:} Maintains an overview of all App Providers, PDOAs, and PDOs.
\end{itemize}
As a user, you first register \emph{once} in a PM-WANI compatible App from an App Provider.
This identity is then used by any PDOA when you connect to their hotspot.
An Internet package is bought from a specific PDOA, and that Internet package can only be used with hotspots aggregated by that specific PDOA.
More precisely, when you first connect to a hotspot behind a given PDOA, that PDOA sees you (via the app) and, if you are a new customer for them, shows you its data packs and registers you as its subscriber once you pay.
To use hotspots of another PDOA, you can still log in using the same app account, but that second PDOA will typically treat you as a new subscriber and ask you to buy one of \emph{its} data packs.
The only exception is when there are roaming agreements between PDOAs, so that each other's subscribers can use any \wifi{} hotspot associated with either PDOA.

\input{table-comparison}

\subsection{Technical details of PM-WANI}

A typical operational flow involves the following three key stages: 

\subsubsection{Hotspot advertisement}

To advertise a new PM-WANI hotspot, and thereby become a PDO, the new hotspot first has to be registered by a PDOA.
This registration includes the hotspot's SSID, MAC address, and location.
The PDOA then registers this metadata of the hotspot with the Central Registry, ensuring that the hotspot becomes discoverable by App Providers.\footnote{At the time of writing, the central registry XML can be found at \url{https://pmwani.gov.in/wani/registry/wani_providers.xml}}

\subsubsection{Hotspot discovery and connection}

To detect nearby PM-WANI hotspots, users must first install a mobile app created by an App Provider.
Such an app, created by an App Provider, scans for nearby APs and queries the central Provider Registry XML to detect if a nearby AP is a PM-WANI hotspot.

Simplified, when a user selects a hotspot, the app generates a \verb|waniapptoken| containing the username, password, timestamp, and the MAC address of the chosen AP~\cite{pm-wani-architecture}.
This token is encrypted with the App Provider’s public key and passed through the Wi-Fi Captive Portal, which re-encrypts it with the PDOA’s private key and forwards it to the App Provider.
The App Provider verifies the identities of both the app and PDOA against the central WANI registry. 
After verification, the captive portal shows the available data packages and payment methods.

Lastly, the specification of PM-WANI also requires that users must be able to log in without an app by manually entering their credentials on the captive portal of the hotspot~\cite{pm-wani-architecture}.

\subsubsection{Payment and monetization}

Before payment, the hotspot grants temporary Internet access for completing any needed payment transactions.

Once payment is confirmed, all devices that authenticate with the same username–password pair share the same session and data pack.
Some PDOAs also employ alternative monetization schemes, e.g., where users can gain Internet access by watching ads~\cite{wani-gavedu}.

\subsection{Security Risks of PM-WANI}

To the best of our knowledge, the security of the PM-WANI architecture has not yet been studied. However, unlike Piso-WiFi, we do not perform a practical attack evaluation against PM-WANI deployments; rather, we conduct a risk assessment based on the architectural design and publicly available specifications of PM-WANI.
Nevertheless, based on our analysis, one major risk is that hotspots are not required to use encryption at the \wifi{} layer, such as WPA2 or WPA3.
This means that PM-WANI hotspots may potentially be vulnerable to the same attacks as covered in Section~\ref{sec:piso:masquerade} and~\ref{sec:piso:rogue} against \piso{}, i.e., they may be vulnerable to user masquerading attacks and rogue-AP-based MitM attacks.

The federated trust model, involving multiple independent PDOs, App Providers, and PDOAs, also introduces additional risks if one of these becomes malicious or compromised. 
For instance, if the central registry is compromised, an adversary can potentially add their own hotspot as a trusted PM-WANI hotspot.
Similarly, if a PDOA is compromised, then an adversary can potentially also add their own hotspot as part of those managed by the PDOA aggregator.

\section{Threat Models and Research Directions \label{sec:threats}}

In this section, we define threat models and propose directions for more secure and cache-enabled \wifi{} hotspots.

\subsection{Threat models}

To design more secure pay-to-use \wifi{} hotspots that can be set up without requiring registration,
we first need concrete threat models.
We propose the following two threat models:

\paragraph{Multi-use model}

In this threat model, we assume the client uses a particular hotspot multiple times.
As a result, we can assume that in the first connection the adversary is not present, but aim to prevent attacks in subsequent connections.

\paragraph{Single-use model}

In this threat model, we assume the client only uses the hotspot once, e.g., when they are traveling.
To provide security, we assume the user is within physical proximity of the hotspot and can for instance see its display.
We assume an adversary will not compromise the physical display or equipment of the hotspot.

\subsection{Securing the \wifi{} connection process}

\paragraph{Multi-use case}

For the multi-use threat model, one solution is a trust-on-first-use protocol, where during the first connection a shared secret is negotiated between the client and hotspot. In later connections, this shared secret, which can also be a public key, is then used to verify the hotspot's authenticity. The main challenge is ensuring backward compatibility and user-friendliness. One solution is to instruct the user to reconnect using an SAE-PK passphrase. This passphrase acts as a digital signature of the hotspot and prevents rogue APs.
Unfortunately, older clients rarely support SAE-PK.
A second option is to force users to reconnect using Enterprise WPA2/3 with a given username and password.
The client can then pin the certificate of the hotspot's RADIUS server to prevent rogue APs, which must be done with care to avoid rogue APs~\cite{appana2025measuring}.
However, it is unclear whether this is user-friendly, i.e., user studies are needed to study feasibility,
or research is needed to see whether this process can be (partly) automated,
e.g., using the provisioning features provided by the Passpoint feature on current devices~\cite{alliance-passpoint-v34}.


In the single-use model, we can rely on the physical proximity to the hotspot.
One option is displaying a QR code with the hotspot's Enterprise certificate~\cite{hasan2025seqr},
though current devices do not yet support scanning such QR codes. 

An alternative is to use the LED lights already present on most \piso{} hotspots as an out-of-band channel.
A client app can record the hotspot’s blink pattern to authenticate session key material exchanged over \wifi{}.
This can be based on the protocol of Saxena et al., who also use blinking LEDs for communication~\cite{saxena2006secure}.

\subsection{Secure local caching solutions}

In addition to providing a secure connection, optimizing the backhaul bandwidth is also essential, especially when there is limited connectivity.
Unfortunately, neither \piso{} nor PM-WANI provide features to locally cache data, e.g., they do not provide an HTTP proxy cache.
Doing so would also be non-trivial because most websites use HTTPS, and a local proxy solution would require breaking the end-to-end security properties of HTTPS~\cite{al2019qos3}.

One solution to balance end-to-end security with caching, without needing to trust the local cache provider, is to separate authentication and encryption of data.
This is feasible since most website content is public and identical for all users~\cite{al2019qos3}, meaning this shared data can be locally cached in plaintext while being authenticated (i.e., signed) by the data's origin.
Initial experiments indicate that a concrete approach to achieve this can be based on Signed Exchanges (SXGs)~\cite{signed-web-exchanges}.
This enables the authentication of web data, independent from where the data is loaded.
The local cache provider can still download data using encryption, but store a plaintext signed SXG object in the cache.
Clients can query this cache, while still being able to verify that the cached data has not been modified.
Interesting future work would be to create a proof-of-concept and to conduct studies to measure the performance advantage of this approach.

\subsection{Other improvements and challenges}

It may also be beneficial to combine the above proposal with letting clients cache data to create a distributed cache, e.g., extensions of~\cite{iyer2002squirrel}.
Other open challenges are ensuring that bandwidth is shared fairly between users, and that users cannot attack each other, i.e., ensuring secure client isolation~\cite{zhou-ndss2026}.

\section{Related Work}
\label{sec:relatedwork}

Several works study deployment strategies for extending Internet connectivity to rural regions.
Gupta provides a description of India’s PM-WANI framework~\cite{gupta2023-pmwani}, including more technical aspects of its design~\cite{gupta2024-satyaspeak}, though without focusing on security.
Kumar et al.\ survey rural Internet connectivity in India and examine technological and economic approaches, including fiber backbones, TV white-space, cellular, and \wifi{}-based access, highlighting that connectivity remains sparse and often unaffordable in many villages~\cite{kumar2022survey}.
Additionally, projects such as TUCAN3G explore small-cell and heterogeneous wireless backhaul architectures tailored to isolated rural communities in developing countries~\cite{martinez2016tucan3g}.
These works primarily focus on coverage, cost, and deployment models, while we focus on the security aspects of low-cost pay-to-use \wifi{} hotspots.

Other works study the security and privacy of public \wifi{} networks and their access mechanisms, for instance,
Ali et al.\ conduct a large-scale measurement study of captive portals in public hotspots and show that many track users and leak personal data to third parties~\cite{ali2019privacy}.
Wang et al.\ analyse the dedicated mini-browsers used by captive portals and found that missing TLS validation, lack of isolation, and other weaknesses enable credential theft and session hijacking~\cite{wang2023capturing}.
James provides an overview of practical attacks against open and poorly configured \wifi{} networks, including MAC-spoofing-based time theft, rogue access points, and man-in-the-middle attacks~\cite{james2021analysis}. 
Finally, Ishtiaq et al.\ perform a security analysis of in-flight \wifi{} paywall systems and uncover design flaws that allow free access and privacy violations~\cite{ishtiaq2025cloud}.

\section{Conclusion and Future Work}
\label{sec:conclusion}

Providing easy-to-use pay-as-you-go access to \wifi{} hotspots in a secure and reliable manner remains a challenge, especially in rural areas.
Due to limitations of the IEEE 802.11 protocol and its implementations, it is currently challenging to design a solution that is both easy to use, secure, and backwards-compatible.
This is evidenced by the practical attacks we confirmed against \piso{} and the PM-WANI system having similar design limitations.

\textit{Future Work.} An open challenge is performing a practical security evaluation of the PM-WANI framework, e.g., determining its practical susceptibility to user masquerading and rogue AP attacks.
Additionally, we believe it is interesting and important future work to design and evaluate improved methods to provide \wifi{} Internet access in rural regions.
A core challenge is doing this in a manner such that the resulting hotspots are compatible with older laptops and smartphones that may not yet support the latest \wifi{} features.
A closely related challenge is also ensuring that, once multiple users are connected, bandwidth is shared fairly between them, and that users cannot attack each other at the network layers above \wifi{}.

\section*{Acknowledgments}

This research is partially funded by the Research Fund KU Leuven and the Cybersecurity Research Programme Flanders.

\bibliographystyle{ACM-Reference-Format}
\bibliography{references}

\end{document}

%% file: table-survey.tex
\begin{table}
    \caption{Summary of the \wifi{} surveys that we performed in India and the Philippines.
    The column Hotspots refers to the number of detected \piso{} and PM-WANI hotspots, respectively.}
    \label{table:survey}
    \centering
    \begin{tabular}{lrrr}
    \toprule
    Country & \#APs & Hotspots & WPA1-TKIP \\
    \midrule
    Philippines & 2\,620 & 125 (4.8\%) & 513 (19.6\%) \\
    India & 6\,464 & 3 (0.05\%) & 851 (13.2\%) \\
    \bottomrule
    \end{tabular}
\end{table}

%% file: fig-piso.tex
\begin{figure}
    \centering
    \includegraphics[width=\linewidth]{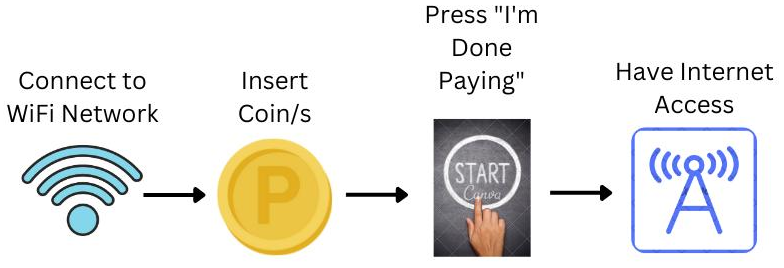}
    \caption{High-level overview of the steps that the user must perform to connect to a \piso{} hotspot.}
    \label{fig:piso}
\end{figure}

%% file: fig-piso-mitm.tex
\begin{figure}
    \centering
    \includegraphics[width=\linewidth]{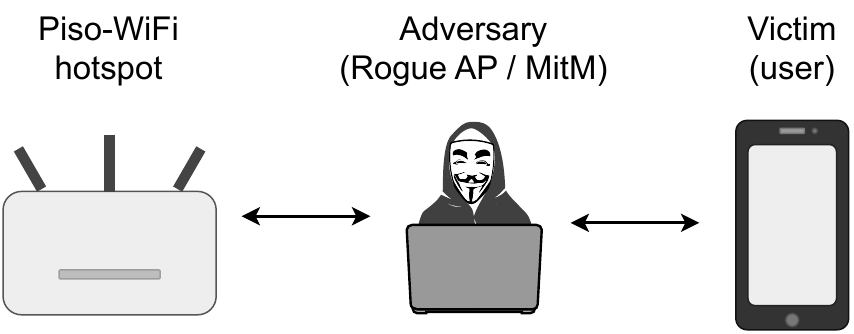}
    \caption{Using a rogue AP to perform a MitM attack against \piso{}.}
    \label{fig:rogue}
\end{figure}

%% file: table-comparison.tex

\begin{table}[t]
\centering
\caption{Comparison between PM-WANI and \piso{}}
\label{tab:pmwani-piso}
\resizebox{\linewidth}{!}{%
\begin{tabular}{@{}p{1.9cm}p{3.1cm}p{3.2cm}@{}}
\toprule
\textbf{Aspect} & \textbf{PM-WANI} & \textbf{\piso{}} \\ 
\midrule
Nature & Federated registry & Local pay-to-use \\ 
Regulation & Government-backed with registration & Informal, unregulated \\ 
Architecture & Multi-entity design & Standalone kiosk \\ 
Payment & With digital wallet or voucher-based & Physical coin payment \\ 
Authentication & App-generated token & Open access with captive pay-to-use portal \\ 
Roaming & Possible using central registry & None; sessions limited to individual kiosks \\ 
\bottomrule
\end{tabular}
}
\end{table}